\documentstyle[multicol,aps,prl,epsf]{revtex}
\begin{document}
\title{Persistence With Partial Survival}
\author{Satya N. Majumdar$^1$ and Alan J. Bray$^2$\\
{\small 1. \em Tata Institute of Fundamental Research, Homi Bhabha Road, 
Mumbai-400005, India}\\
{\small 2. \em Department of Theoretical Physics, The University, Manchester,
M13 9PL, United Kingdom}
}
\date{\today}
\maketitle

\begin{abstract}
We introduce a parameter $p$ called partial survival in the persistence of
stochastic processes and show that for smooth processes the persistence
exponent $\theta(p)$ changes continuously with $p$, $\theta(0)$ being 
the usual persistence exponent. We compute $\theta(p)$ exactly for
a one-dimensional deterministic coarsening model, and approximately
for the diffusion equation. Finally we develop an exact, systematic  
series expansion for $\theta(p)$, in powers of $\epsilon=1-p$, for a  
general Gaussian process with finite density of zero crossings.

\end{abstract}

\begin{multicols}{2}
Recently considerable theoretical and experimental effort has been devoted
to understanding first-passage statistics in nonequilibrium systems.
These include the Ising, Potts and time-dependent deterministic Ginzburg-Landau
(TDGL) models undergoing zero-temperature phase-ordering
dynamics\cite{derrida1,derrida2,bray,majsire}, the diffusion 
equation with random initial conditions\cite{diffusion,watson}, the global
magnetization undergoing critical dynamics\cite{crit}, several
reaction-diffusion systems\cite{cardy}, fluctuating
interfaces\cite{krug} and a randomly accelerated particle\cite{burkhardt}.
Typically one is interested in `persistence', i.e.\ the
probability $P_0(t)$ that, at a fixed point in space, the stochastic
process (such an Ising spin or the diffusion field) does
not change sign up to time $t$. In the examples mentioned above, this
probability decays as a power for large time $t$, $P_0(t)\sim
t^{-\theta}$, where the persistence exponent $\theta$ is nontrivial
due to the non-Markovian nature of the process in time {\it {at a fixed
point in space}}. This exponent has recently been measured experimentally
in a $2$-d liquid crystal system\cite{yurke}, and also for $2$-d soap
froth\cite{tam} and breath figures \cite{martin}. The theoretical 
computation of $\theta$ however, despite a few exact and approximate 
results, remains a major challenge. 

Even for the simple diffusion equation, $\partial_{t}\phi={\nabla}^2\phi$
starting from random initial configuration, the exponent $\theta$ is
known only numerically and within an independent interval approximation
(IIA)\cite{diffusion}, though there is a recent conjecture\cite{newman} 
for an exact $\theta$ that remains to be proved. The IIA result, though in 
excellent agreement with numerical simulations, is hard to improve 
systematically. The central result of this Letter is to derive a systematic 
series expansion for $\theta$ in terms of a suitable expansion parameter. 
This expansion is {\em {exact}} order by order and when truncated at second 
order already gives good results for the diffusion equation. 
But this exact series expansion technique is more general and goes beyond 
the diffusion equation. We show that it can be applied to compute
the persistence exponent, order by order, for a wide class of stochastic
processes which includes the diffusion equation, random acceleration and the 
$1$-d TDGL model as special cases.

Our result is also useful for a related problem which has wide
applications in diverse fields ranging from information theory to
stock markets and oceanography. Consider a stochastic Gaussian {\em
{stationary}} process $X(T)$ characterized completely by its two-time
correlator, $\langle X(0)X(T)\rangle=f(T)$. The process $X(T)$ can be
used to model, e.g., the current in an electrical circuit or the price
of a stock. Given $f(T)$, what is the probability $P_0(T)$ that the signal
stays above (or below) a certain level , say zero, up to time $T$? This 
problem has been studied for many years\cite{slepian,reviews} and it is
known that if $|f(T)|<1/T$ for large $T$, then $P_0(T)\sim \exp (-\theta
T)$ for large $T$\cite{slepian}. The exponent $\theta$ depends
quite sensitively on the full function $f(T)$ and is very hard to
compute for general $f(T)$\cite{majsire}. For a Markov process, where
$f(T)=\exp(-\lambda T)$ for all $T$, it is known that
$\theta=\lambda$\cite{slepian}. However for non-Markov processes, where 
$f(T)$ is not a pure exponential, very little is known. Only 
recently a perturbation theory result for $\theta$ was developed for 
processes close to Markovian\cite{majsire,klaus}. 

The persistence problem for the diffusion equation in $d$-dimensions 
can be exactly mapped to a Gaussian stationary process, with 
$f(T)=[{\rm sech}(T/2)]^{d/2}$, by identifying $T=\ln t$ and 
$X(T)=\phi(x,t)/{\sqrt {\langle\phi^2(x,t)\rangle}}$\cite{diffusion}. 
The probability of no zero crossing then decays as 
$P_0(T)\sim \exp(-\theta T)= t^{-\theta}$. The series expansion 
technique that we develop below can be used to compute $\theta$ for 
arbitrary $f(T)$ as long as $f(T)\sim 1-aT^2+\ldots$ for small $T$. 
Such Gaussian processes are called {\em {smooth}} as they have a finite 
density of zero crossings, $\rho ={\sqrt {-f''(0)}}/{\pi}$\cite{rice}.

The key strategy underlying our technique is to first generalize the
usual persistence problem by introducing a partial survival factor $p$
as follows. The usual persistence, say in the diffusion equation, is the
fraction of points in space where the diffusion field has not changed
sign even once up to time $t$. One way to compute this is to start with a
random initial configuration of the field and put a particle at
each point in space to act as a counter. At subsequent times, whenever the  
field changes sign at any point, the particle there dies. The persistence
is simply the fraction of particles still surviving at time $t$. 
We now generalize this by assigning the rule that whenever the field 
changes sign at a point, the particle there survives with probability $p$ 
and dies with probability $(1-p)$. We then compute the fraction of particles, 
$P(p,t)$ left after time $t$. Thus, $p=0$ corresponds to usual persistence
$P(0,t)$. A somewhat similar generalization was recently studied 
in the context of ``adaptive persistence" problems\cite{majcornell}. 

This generalization has several implications. It is easy to see that if
$P_n(t)$ denotes the probability of $n$ zero crossings in time $t$ of the
underlying single site process, then $P(p,t)$ is simply the generating
function,
\begin{equation}
P(p,t)= \sum_{n=0}^{\infty} p^n P_n(t).
\label{eq:genfunc}
\end{equation}
For $p=0$, $P(0,t)=P_0(t)$, the usual persistence, decaying for large
$t$ as $t^{-\theta(0)}$. In the other limit, $p=1$, the
particles always survive: $P(1,t)=1$, implying $\theta(1)=0$.
It is interesting to analytically continue Eq.\ (\ref{eq:genfunc}) 
to negative $p$. In fact, for $p=-1$ this is simply the
autocorrelation function, $P(-1,t)=A(t)=\langle \rm{sgn}(\phi(x,0))\,
\rm{sgn}(\phi(x,t)\rangle$, which decays as $t^{-\lambda/2}$, where 
$\lambda$ is a well-studied exponent in phase-ordering systems\cite{fisher}. 
In fact, we show below that for smooth processes, 
$P(p,t)\sim t^{-\theta(p)}$ for large $t$ where the exponent $\theta(p)$
depends continuously on $p$ as $p$ varies from $-1$ to $+1$. 
Moreover, the quantity $A_p(t)= P(-p,t)/P(p,t)$ is just the 
autocorrelation function averaged only over points with surviving particles, 
when the survival probability is $p$. So if $A_p(t)\sim t^{-\lambda_p}$, we 
have $\lambda_p = \theta(-p)-\theta(p)$. this generalization thus puts
both the autocorrelation and the persistence exponents as members of a
wider family of exponents.

We first establish the continuous dependence of $\theta(p)$ on $p$ for
smooth processes by computing $\theta(p)$ exactly for a non-Gaussian
process, namely the $1$-d deterministic TDGL model, and then approximately
within IIA for the diffusion equation. We then proceed to compute
$\theta(p)$ for any smooth Gaussian stationary process by
expanding around $p=1$. This series expansion result for $\theta(p)$ in
powers of $\epsilon=1-p$ is exact order by order.     

If a system, such as the Ising model, is quenched from a high-temperature
disordered phase to zero temperature, domains of `up' and `down' phases
form and grow with time. The evolution of the order-parameter field $\phi$
can be modelled by the deterministic TDGL equation,
$\partial_t\phi={\nabla}^2 \phi-V'(\phi)$, where $V(\phi)$ is a symmetric
double well potential with minima at $\phi=\pm 1$. In $1$-d, at late times
the system breaks up into alternate `up' and `down' domains and coarsens
by successively eliminating the boundaries of the smallest domain, i.e., 
by flipping the signs of $\phi$ simultaneously at all points
inside the smallest domain\cite{nagai}. The density of persistent, or `dry'
parts where $\phi$ has not changed sign then scales as $\sim {\langle
l\rangle}^{-\theta(0)}$ where $\langle l\rangle$ is the average length
of growing domains, which serves as `time' in this problem. The exponent
$\theta(0)$ was computed exactly by noting that the dynamics does 
not generate correlations between neighboring domains\cite{bray}. We now
introduce the partial survival factor $p$ in this dynamics. 

We start with a random distribution of intervals or domains and assign
a particle to each point in space. The dynamics merges the smallest interval 
$I_{min}$ with its two neighbors
$I_1$ and $I_2$ to make one single interval $I$. The lengths $l(I)$
and the `dry' part $d(I)$ (i.e, the number of live particles in the
interval $I$) evolve as, $l(I)=l(I_1)+l(I_2)+l(I_{min})$ and
$d(I)=d(I_1)+d(I_2)+pd(I_{min})$. Thus the only difference from the
calculation in Ref.\ \cite{bray} is the $p$-dependent term in the dry part.
The rest of the calculation is similar to that in Ref.\ \cite{bray} and we
just outline the method without details. One writes down the evolution
equations for the number of intervals of length $l$ and the average dry
part carried by such an interval, and one solves exactly for the associated 
scaling functions by taking Laplace transforms. Demanding that
the first moments of these scaling functions are finite gives a
transcendental equation for $\theta(p)$,
\begin{eqnarray}
\int_{0}^{\infty}dt && e^{-t}\,t^{-1-\theta}\bigl[(1-p)(1-t-e^{-t})e^{r(t)} 
\nonumber \\
&& + 2{\theta}(1+p)t+{\theta}(1-p)t^2 e^{-r(t)}\bigr ]=0,
\label{eq:trans}
\end{eqnarray}
where $r(t)=-\gamma-\sum_{n=1}^{\infty}(-t)^n/n\,n!$, $\gamma$
being Euler's constant. Clearly, for $p=1$ one gets $\theta(1)=0$ from
the above equation as expected. For $p=0$, it reduces to the equation for 
$\theta(0)$ as obtained in Ref.\ \cite{bray}. For $p=-1$, one recovers the 
equation for $\theta(-1)=\lambda$ of Ref.\ \cite{bray2}. Fig.\ 1 shows 
$\theta(p)$ as a function of $p$ for $-1\le p\le 1$, obtained by numerically 
solving Eq.\ (\ref{eq:trans}).

\begin{figure}
\narrowtext
\centerline{\epsfxsize\columnwidth\epsfbox{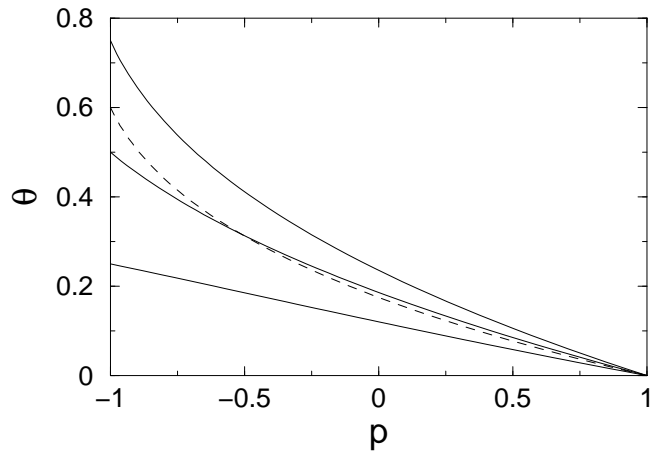}}
\caption{Dashed line: The exponent $\theta(p)$ for the $1$-d TDGL 
model, obtained from Eq.\ (\ref{eq:trans}). Solid lines: The IIA estimates 
for $\theta(p)$ for the diffusion equation in (bottom to top) $1$, $2$ and 
$3$ dimensions.}
\label{theta}
\end{figure}

We now turn to the diffusion equation, $\partial_t\phi=\nabla^2\phi$,
starting from a random initial configuration. We first carry out a
numerical simulation to compute $P(p,t)$ for finite $p$ following
the procedures of Ref.\ \cite{diffusion}. Fig.\ 2 shows the asymptotic
decay of $P(p,t)$ with $t$ on a log-log plot for $p=0$ and $p=0.5$ in 
$1$-d. Clearly the exponents are quite different. For example, for $p=0$,
$\theta(0)=0.1207\pm 0.0005$ as in Ref.\ \cite{diffusion} but for $p=0.5$, 
we find $\theta(1/2)=0.0588\pm 0.0005$. Unfortunately we have not been able 
to compute $\theta(p)$ exactly. However, the IIA used in 
Ref.\ \cite{diffusion} to compute $\theta(0)$ can be easily extended to 
compute $\theta(p)$ very accurately for all $p$ as we now show. 

\begin{figure}
\narrowtext
\centerline{\epsfxsize\columnwidth\epsfbox{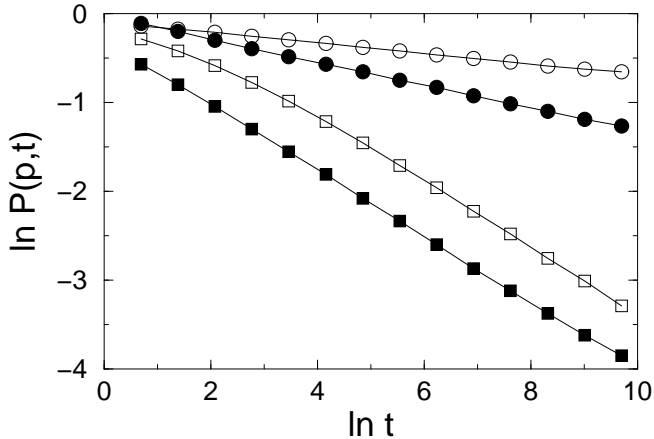}}
\caption{Time-dependence of the generalized persistence probability, $P(p,t)$, 
from numerical simulations. Circles: the $1$-d diffusion equation for $p=0$ 
(filled) and $p=0.5$ (open). Squares: the $1$-d Glauber model at 
zero temperature for $p=0$ (filled) and $p=0.5$ (open). The lines are guides 
to the eye.}
\label{satya}
\end{figure}

Consider the normalized process $X(T)=\phi/{\langle
{\phi^2}\rangle}$ as a function of $T=\ln t$. The zero-crossing events of 
$\phi$ are the same as those of $X$, but $X(T)$ is a {\em {stationary}} 
Gaussian process characterized completely by its two-time correlator, 
$f(T)=\langle X(0)X(T)\rangle=[\rm{sech}(T/2)]^{d/2}$. With a non-zero 
survival factor $p$, the fraction of live particles after time $T$ is then,
$P(p,T)=\sum_{n=0}^{\infty}p^n P_n(T)$ and will decay at late times as 
$\exp{(-\theta(p)T)} = t^{-\theta(p)}$. To evaluate $P_n(T)$, the 
probability of $n$ zero crossings by $X$ in time $T$, we note that the 
Laplace transforms, $\tilde{P}_n(s)=\int_0^{\infty} dT \exp (-sT)P_n(T)$, 
were evaluated in Ref.\ \cite{diffusion} using IIA, i.e., assuming that 
successive intervals between zero crossings of $X$ are statistically
independent. Using these results from Ref.\ \cite{diffusion}, and carrying
out the sum over $n$, gives $\tilde{P}(p,s)$. Since $P(p,T)\sim
\exp(-\theta(p)T)$ for large $T$, $P(p,s)$ will have a simple pole
at $s=-\theta(p)$. Using this, we finally get $\theta(p)$ as
a solution of the equation,
\begin{eqnarray}
\frac{1-p}{1+p} &=& \theta\pi\sqrt{\frac{2}{d}}\left\{1 + 
\frac{2\theta}{\pi}\int_0^{\infty}dT\,\exp(\theta T)\,\right. \nonumber \\
& & \hspace*{1.5cm}\times \left. \sin^{-1}[\rm{sech}^{d/2}(T/2)]\right\}.
\label{eq:pole}
\end{eqnarray}
The solution is plotted in Fig.\ 1 for $d=1,2,3$. We note in the two
extreme limits, $p=1$ and $p=-1$, the IIA gives $\theta(1)=0$
and $\theta(-1)=d/4$, which are exact. For intermediate values of $p$, 
the IIA results are in excellent agreement with numerical simulations.
For example, for $p=1/2$ the IIA gives $\theta_{IIA}=0.05823044\ldots$, 
compared to $\theta_{sim}=0.0588\pm .0005$.

Having established the continuous $p$-dependence of $\theta(p)$ for two
smooth processes, we now derive an exact series expansion of $\theta(p)$ 
near $p=1$ for a general smooth, Gaussian, stationary process $X(T)$, 
characterized by its two-time correlator $f(T)$. The
basic idea is straightforward. We start with the definition 
(\ref{eq:genfunc}) of $P(p,t)$ as a generating function. Writing
$p^n=\exp(n\ln p)$ and expanding the exponential, we obtain an expansion
in terms of the moments of $n$, the number of zero crossings: 
\begin{equation}
\ln P(p,T)= \sum_{r=1}^{\infty}{{(\ln p)^r}\over {r!}}{\langle
n^r\rangle}_c,
\label{eq:moment}
\end{equation}
where ${\langle n^r\rangle}_c$ are the cumulants of the moments. Using
$p=1-\epsilon$, we express the right-hand side as a series in powers of 
$\epsilon$. Since $P(p,T)$ is expected to decay for large $T$ as $\exp
(-\theta(p)T)$, we obtain a series expansion of $\theta(p)$
by taking the limit,  
\begin{equation}
\theta(p)= -\lim_{T\to \infty} {{1}\over {T}}\ln P(p,T) 
= \sum_{r=1}^{\infty}a_r{\epsilon}^r.
\end{equation}
The coefficients $a_r$'s involve the cumulants.

Fortunately the computation of the moments of $n$ is relatively 
straightforward, though tedious for higher moments. For example, the first 
moment $\langle n\rangle$, i.e.\ the expected number of zero crossings in 
time $T$, was computed by Rice\cite{rice}: 
$\langle n\rangle=T\sqrt {-f''(0)}/{\pi}$, implying 
$a_1=\sqrt{-f''(0)}/{\pi}$. The second moment, $\langle n^2\rangle$, was 
computed by Bendat\cite{bendat}. Using this result
and after some algebra we have computed the coefficient $a_2$, which 
already looks complicated. We just quote the final result here (details 
will be published elsewhere\cite{majbray}): 
\begin{equation}
a_2 = \frac{1}{\pi^2} \int_0^{\infty} [S(\infty)-S(T)]dT,
\end{equation}
where $S(T)$ is given by
\begin{equation}
S(T)={ {\sqrt {M_{22}^2-M_{24}^2} }\over {[1-f^2(T)]^{3/2} } } \bigl [1+
H{\tan}^{-1}H\bigr ],
\end{equation}
with $H=M_{24}/{\sqrt {M_{22}^2-M_{24}^2} }$. The $M_{ij}$'s are the
cofactors of the $4\times4$ symmetric correlation matrix $C$ between $4$
Gaussian variables $[X(0), {\dot {X}}(0), X(T), {\dot {X}}(T)]$. The
elements of $C$ can easily be computed from the correlator $f(T)$. For
example, $C_{11}=\langle X(0)X(0)\rangle=f(0)$, 
$C_{14}=\langle X(0){\dot {X}}(T)\rangle=f'(T)$,
$C_{24}=\langle {\dot {X}}(0){\dot {X}}(T)\rangle=-f''(T)$ and so on.

Although these expressions look complicated, in many cases the
function $S(T)$ can be evaluated explicitly and the integral for 
$a_2$ can be performed analytically. For example, for
$2$-d diffusion equation, where $f(T)= \rm{sech}(T/2)$, we get
\begin{equation}
\theta(p=1-\epsilon)= {1\over {2\pi}}\epsilon + ({1\over {\pi^2}}-{1\over
{4\pi}}){\epsilon}^2 + O({\epsilon}^3).
\end{equation}    
Keeping terms up to second order and putting $\epsilon=1$ (in the same
spirit as $\epsilon$ expansion in critical phenomena) gives
$\theta(0)=(\pi+4)/{4{\pi}^2}=0.180899...$, just $3.5\%$
below the simulation value, $\theta_{sim}=0.1875\pm 0.0010$\cite{diffusion}.
Note that though the IIA estimate, $\theta_{IIA}=0.1862$ is even closer 
to the simulation, it can not be improved systematically.
The series expansion estimate, on the other hand, can be improved
systematically order by order.

We have also computed, for the first time, the third moment $\langle
n^3\rangle$ for a general smooth correlator $f(T)$. We then use
this to compute $a_3$. The expressions involve
the elements of a $6\times 6$ correlation matrix and are not particularly
illuminating\cite{majbray}, so we skip the details
here. As an example, we computed the series up to third order for
the random acceleration process, $d^2x/dt^2=\eta$ 
($\eta$ is a Gaussian white noise) which can be transformed to a
Gaussian stationary process with
$f(T)=[3\exp(-T/2)-\exp(-3T/2)]/2$\cite{diffusion}. 
We find
\begin{equation}
\theta(p)= {{\sqrt 3}\over {2\pi}}\left(\epsilon - {1\over 6}
{\epsilon}^2 +{11\over 72}\epsilon^3 +O(\epsilon^4)\right).
\end{equation} 
Putting $\epsilon=1$, we get to third order, $\theta(0)=0.271835775...$
which should be compared to its exact value $0.25$\cite{burkhardt}. We
note that the series oscillates around the exact value $0.25$ as the order
increases.

We note that the series expansion will fail for non-smooth Gaussian 
processes, whose moments of zero crossings are not finite. As an example,
consider ordinary Brownian motion, $dx/dt=\eta$, which can be mapped to a 
stationary Gaussian Markov process with correlator $f(T)=\exp(-T/2)$ using
the change of variables discussed before. For this process it is well
known\cite{slepian} that the moments of zero crossings are infinite: 
if the process crosses zero once, then it crosses again infinitely 
many times immediately afterwards\cite{barbe}. Thus only the $n=0$
term contributes to the sum (\ref{eq:genfunc}), giving 
$P(p,T)\approx P(0,T)\sim \exp (-T/2) =  t^{-1/2}$ for large $t$.
Thus $\theta(p)=1/2$ for all $0\le p <1$, except at $p=1$
where $\theta(1)=0$. Since $\theta(p)$ is discontinuous at $p=1$,
no expansion around $p=1$ is possible. 

The same conclusion holds for the $T=0$ Glauber dynamics of the Ising model. 
In this case, the usual persistence exponent $\theta(0)$ was
recently computed exactly in $1$-d\cite{derrida2} and approximately in 
higher dimensions\cite{majsire}. The exact value in $1$-d is
$\theta(0)=3/8$\cite{derrida2}. Even though the spin $S_i(t)$ at a given 
site $i$ is no longer a Gaussian process, it is non-smooth nevertheless, 
i.e.\ if a spin flips once, it usually flips many times immediately
afterwards. 
This fact can be tested easily by computing the exponent
$\theta(p)$ for nonzero $p$. In Fig.\ 2 we show the asymptotic
dependence of $P(p,t)$ on $t$ on a log-log plot for $p=0$ and $p=1/2$
for the $1$-d $T=0$ Glauber model. In contrast to the diffusion
case, the asymptotic slopes are the same and given by $0.375\pm 0.002$.
We have checked this fact for other values of $p$, and conclude that
$\theta(p)$ is independent of $p$ for $0 \le p <1$ \cite{krapivsky}, 
while clearly $\theta(1)=0$. Thus the $p$ dependence of $\theta(p)$ 
provides important information about the nature of the smoothness of the 
underlying stochastic process. 

We thank Deepak Dhar for useful discussions.

\end{multicols}


\begin{references}
\bibitem{krapivsky} P. L. Krapivsky (private communication) has shown that, 
for the 1-d Glauber model, the result $\theta(p) = 3/8$ for $|p|<1$ follows 
from the scaling form for $P_n(t)$ proposed by E. Ben-Naim, L. Frachebourg 
and P. L. Krapivsky, Phys.\ Rev.\ E {\bf 53}, 3078 (1996).
\end{references}
\end{document}